\expandafter\def\csname @nil\endcsname{0=0}
\documentclass{aa}  
\usepackage{booktabs}
\usepackage{tablefootnote}
\usepackage{longtable}
\usepackage{placeins}
\usepackage{lscape}
\usepackage{mathrsfs}
\usepackage{hyperref}
\hypersetup{
  colorlinks=true,
  linkcolor=blue,
  citecolor=blue,
  urlcolor=blue
}
\usepackage[detect-weight=true, mode=text]{siunitx}

\usepackage{txfonts}
\usepackage{natbib}
\usepackage{amssymb}
\bibpunct{(}{)}{;}{a}{}{,}

\begin{document}  

\title{Unveiling the properties of pulsating low-mass helium-core white dwarfs through TESS asteroseismology I. First results
}

\author{Leila M. Calcaferro\inst{1,2}, Nikoo Hosseininezhad\inst{3}, Murat Uzundag\inst{4}, Alejandro H. Córsico\inst{1,2}, Keaton J. Bell\inst{3,5}, Leandro G. Althaus\inst{6}, \and J.J. Hermes\inst{7}} 
\offprints{lcalcaferro@fcaglp.unlp.edu.ar} 

\institute{$^1$ Grupo  de Evoluci\'on  Estelar y  Pulsaciones,  Facultad de 
           Ciencias Astron\'omicas  y Geof\'{\i}sicas, Universidad
           Nacional de La Plata, Paseo del Bosque s/n, (1900) La
           Plata, Argentina\\   
           $^{2}$ Instituto de Astrof\'{\i}sica
           La Plata, CONICET-UNLP, Paseo  del Bosque s/n, (1900) La
           Plata,
           Argentina\\ 
           $^{3}$ City University of New York Graduate Center, 365 5th Ave, New York, NY 10016, USA\\           
           $^{4}$ Institute of Astronomy, KU Leuven, Celestijnenlaan 200D, 3001, Leuven, Belgium \\
           $^{5}$ Physics Department, Queens College, City University of New York, 6530 Kissena Blvd, Queens, NY 11367, USA\\
           $^{6}$ Departament de Física, Universitat Politècnica de Catalunya, c/ Esteve Terradas 5, 08860 Castelldefels, Spain\\
           $^{7}$ Department of Astronomy, Boston University, 725 Commonwealth Ave., Boston, MA 02215, USA\\
\email{lcalcaferro@fcaglp.unlp.edu.ar}}
\date{Received, }  

\abstract {Recent space-based photometry, particularly from the TESS mission, has transformed the asteroseismological study of pulsating white dwarfs (WDs). In particular, it has opened new possibilities for probing the internal structure of low-mass (LM) helium (He)-core WDs.}
{We present a detailed asteroseismological analysis of six pulsating LM WD stars, including new and updated TESS photometry analyzed homogeneously.}
{We processed short- and ultra-short-cadence TESS observations of TIC~290904838 (J1112), TIC~156064657, TIC~33717565, TIC~344130696, TIC~72637474, and TIC~188087204, and analyzed the resulting pulsation spectra. We then carried out a detailed asteroseismological analysis using fully evolutionary models of LM He-core WDs that allow for varying hydrogen (H)-envelope thicknesses. We also estimated spectroscopic/photometric stellar masses when atmospheric parameters are available.}
{We report first TESS-based frequencies for J1112 and provide revised or expanded frequency solutions for the remaining targets. The asteroseismological analysis yields relatively well-constrained solutions for three stars, a representative but more tentative solution for one target, and constrained ranges for the remaining two. The inferred solutions span a broad range of H-envelope thicknesses, although some of the asteroseismological inferences remain tentative because of the limited number of observed periods available for the analysis. For most objects, the derived spectroscopic/photometric stellar masses are broadly compatible with the asteroseismological values.}
{This is the first homogeneous TESS-based asteroseismological study of a small sample of pulsating LM WDs. 
Our results suggest that 
LM WDs can harbor H envelopes with a range of thicknesses, 
from canonical (thick) to very thin, as in average-mass H-rich pulsating WDs. They also provide a useful reference point for future studies of larger samples, which will hopefully benefit from richer mode sets and improved mode identification.}
 
  \authorrunning{Calcaferro et al.}
  \titlerunning{Asteroseismology of ELMVs with TESS data}
  \maketitle

\section{Introduction}  
\label{intro}  

White dwarf (WD) stars represent the ultimate fate for the vast majority of stars in the Universe, including our Sun \citep{2008ARA&A..46..157W,2008PASP..120.1043F,2010A&ARv..18..471A,2022PhR...988....1S}.  These remnants are the expected outcomes for progenitor stars with masses smaller than $8 - 10.5\ M_{\sun}$, depending on the initial metallicity \citep[e.g.,][]{2014MNRAS.441..582D}. Among their many properties, the stellar mass of WDs plays a vital role in astrophysical research, since it constrains the initial-to-final mass relation and is a key ingredient in deriving the WD luminosity function \citep[see e.g.,][]{2015MNRAS.448.1779B,2016NewAR..72....1G,2017ApJ...837..162K,2018ApJ...860L..17E,2019ApJ...871L..18C}. Most WDs have hydrogen (H)-rich atmospheres, and are therefore classified as DA WDs, with an average mass of about $0.6\ M_{\sun}$ and likely carbon-oxygen (CO) cores \citep{2016MNRAS.455.3413K,2016MNRAS.461.2100T}. At the higher end of the mass distribution ($\gtrsim  1.05\ M_{\sun}$), ultra-massive WDs may instead harbor CO or oxygen-neon (ONe) cores \citep[see e.g.,][]{2021A&A...646A..30A,2021A&A...649L...7C,2022MNRAS.516L...1C,2021ApJ...906...53S,2025ApJ...994..255J}. Conversely, at the low-mass end, helium (He)-core WD stars are characterized by $M_{\star} \lesssim 0.45\ M_{\sun}$ and are often referred to as low-mass (LM) WDs\footnote{Note that LM WDs with masses higher than $\sim 0.33\ M_{\sun}$ may also harbor CO cores \citep[see e.g.,][]{2000MNRAS.319..215H,2009A&A...507.1575P}.} \citep{2010A&ARv..18..471A}. 

The formation of LM WDs typically results from binary evolution, occurring through stable mass transfer via Roche-lobe overflow
 in close binary systems or via unstable mass loss during common-envelope phases \citep{2013A&A...557A..19A,2025A&A...699A.280A,2016A&A...595A..35I,2018ApJ...858...14S,2019ApJ...871..148L}. In both channels, the envelope of the red-giant progenitor is removed before the He flash, so He ignition is avoided and the remnant retains a He core \citep{2013A&A...557A..19A,2016A&A...595A..35I}. Binary evolution is required to explain the lowest-mass systems ($M_{\star} \lesssim 0.3\ M_{\sun}$), since the Universe is not old enough to produce such WDs through single-star evolution. This is particularly true for extremely low-mass (ELM) WDs, often defined as those with $M_{\star} \lesssim 0.18 - 0.20\ M_{\sun}$. While some authors adopt $\sim 0.3\ M_{\sun}$ as the upper mass limit for ELM WDs \citep[see e.g.,][]{2016ApJ...818..155B}, here we follow \cite{2014A&A...569A.106C} \citep[see also][]{2019A&ARv..27....7C} and define an ELM WD as one whose progenitor did not undergo H-shell flashes. This criterion \citep[which depends on metallicity; see e.g.,][]{2002MNRAS.337.1091S,2016A&A...595A..35I}, is physically motivated and crucial because the presence or absence of H flashes affects the evolutionary timescales and pulsational properties.

Studying the evolution of LM and ELM WDs is pivotal not only for constraining binary interactions but also for investigating their broader impact across various astrophysical contexts.
In compact binary systems, LM WDs are expected to play a key role as sources of electromagnetic and gravitational waves, especially for space-based detectors such as the upcoming Laser Interferometer Space Antenna \citep[LISA][]{2017arXiv170200786A}. Some of these binaries may also represent potential progenitors of Type Ia supernovae \citep{1984ApJ...277..355W,1984ApJS...54..335I,2007ApJ...662L..95B}.

The number of LM and ELM WDs has steadily increased thanks to large spectroscopic efforts, particularly the ELM Survey, a spectroscopic program designed to identify LM WD binaries using photometry from large sky surveys \citep{2010ApJ...723.1072B,2012ApJ...744..142B,2013ApJ...769...66B,2016ApJ...818..155B,2020ApJ...889...49B,2022ApJ...933...94B,2011ApJ...727....3K,2012ApJ...751..141K,2015ApJ...812..167G}. \cite{2020ApJ...894...53K} extended the ELM Survey to the southern sky and demonstrated that a Gaia-based approach is effective in detecting ELM WD binaries \citep[see also][]{2023ApJ...950..141K}. 

Multi-periodic brightness variations likely produced by global stellar pulsations have been 
detected in several LM and ELM WDs, defining the class of variable WDs known as ELMVs. To date, more than 20 candidates have been reported \citep{2012ApJ...750L..28H,2013ApJ...765..102H,2013MNRAS.436.3573H,2015MNRAS.446L..26K,2017ApJ...835..180B,2018A&A...617A...6B,2018MNRAS.478..867P,2020NatAs...4..690P,2021ApJ...922..220L,2021ApJ...912..125G,2022MNRAS.511.1574R,2023ApJ...950..141K,2023ApJ...958..101B,2024A&A...685A...9A}. The advent of space photometry, and in particular observations from the Transiting Exoplanet Survey Satellite \citep[TESS;][]{2015JATIS...1a4003R}, has expanded the sample of ELMVs and provided new observations of previously known ones \citep{2021ApJ...922..220L,2022MNRAS.511.1574R,2025ApJ...984..112R,2024A&A...684A..76B}.  Photometric variations have also been detected in stars considered to be probable precursors of LM WDs \citep[the pre-ELMV variable stars; e.g.,][]{2013Natur.498..463M,2014MNRAS.444..208M,2016ApJ...822L..27G,2020ApJ...888...49W,2022ApJ...936....5W,2022MNRAS.515.4702L}. The existence of ELMVs and pre-ELMVs provides us with a unique opportunity to probe the interiors of these stars and test their formation scenarios through asteroseismology \citep{2008ARA&A..46..157W,2010A&ARv..18..471A,2019A&ARv..27....7C, 2026enap....3...75B}. 

The brightness variations observed in ELMVs are consistent with gravity ($g$) modes, likely driven by a combination of the $\kappa-\gamma$ \citep{1989nos..book.....U} and convective driving \citep{1991MNRAS.251..673B} mechanisms, both occurring in the partial ionization region of H
\citep{2010ApJ...718..441S,2012A&A...541A..42C,2013ApJ...762...57V,2016A&A...585A...1C}. Detailed nonadiabatic pulsation computations predict that short-period $g$ modes can also be destabilized by the $\epsilon$ mechanism due to stable H burning, particularly for models with $M_{\star} \lesssim 0.18\ M_{\sun}$ on their final cooling tracks \citep{2014ApJ...793L..17C}, or before the onset of the CNO flashes, on their early WD cooling branches, in LM WDs with $M_{\star} > 0.18\ M_{\sun}$ \citep{2021A&A...647A.140C}.  

Asteroseismology of WDs provides stellar parameters by matching observed pulsation periods with theoretical periods from stellar models \citep[e.g.,][]{2026enap....3...75B}. Fully evolutionary models, evolved self-consistently from the zero-age main sequence (ZAMS), provide a robust framework for this purpose and have been extensively applied by the La Plata Group\footnote{\url{http://evolgroup.fcaglp.unlp.edu.ar/publications.html}} to several classes of pulsating WDs \citep[see e.g.,][]{2007A&A...461.1095C,2007A&A...475..619C,2008A&A...478..869C,2012A&A...541A..42C,2012MNRAS.420.1462R,2017A&A...607A..33C,2018A&A...620A.196C,2024A&A...686A.140C,2022MNRAS.513.2285U,2023MNRAS.526.2846U}. This approach allows the chemical structure and internal stratification to be determined in a fully consistent way. 
Alternative approaches employ static parametric WD structures  \citep[e.g.,][]{2018Natur.554...73G} or parameterized evolutionary models \citep[e.g.,][]{2019ApJ...871...13B}. For ELMVs, fully evolutionary models tailored to LM and ELM WDs have been successfully used to probe their internal structure and evolutionary status \citep{2017A&A...607A..33C,2018A&A...620A.196C,2018MNRAS.479.1267K}. Other asteroseismological analyses have also been conducted for individual pulsating LM WDs employing the mentioned alternative approaches \citep[see e.g.,][]{2023ApJ...943..113S,2023ApJ...958..101B}. 

Most evolutionary calculations of LM and ELM WDs are based on progenitor stars that experienced stable mass transfer \citep{2013A&A...557A..19A,2016A&A...595A..35I}, which produce stellar models with thick (canonical) H envelopes that sustain residual stable H burning and thus have extremely long cooling times even at high effective temperatures \citep{2013A&A...557A..19A}. However,  stars with thin H envelopes --- unable to sustain residual H burning --- are also plausible, as observations suggest \citep[see e.g.,][]{2009ApJ...699...40S,2021A&A...650A.102I}, and would cool to much lower temperatures than their thick-envelope counterparts \citep[e.g. $2500\ $K vs. $7000\ $K;][]{2018A&A...614A..49C}.  Motivated by this diversity, \cite{2018A&A...620A.196C} explored a range of H-envelope masses ($M_{\rm H}$) and showed, through asteroseismology, that some ELMVs are better reproduced with thinner-than-canonical H envelopes. Notably, reducing $M_{\rm H}$ increases $\log g$ for a given $M_{\star}$, thereby biasing mass estimates derived from $(T_{\rm eff},\log g)$ if the possible range of envelope thicknesses is not taken into account \citep{2018A&A...614A..49C,2025A&A...699A.280A}. Hence, stellar masses should be inferred using evolutionary tracks that sample the relevant range of $M_{\rm H}$ to avoid systematic offsets \citep[see][]{2024A&A...691A.194C}.

In this work, we present an updated and homogeneous analysis of TESS  photometry and asteroseismology for six ELMV stars: TIC~290904838 (J1112), TIC~156064657, TIC~33717565, TIC~344130696, TIC~72637474, and TIC~188087204. For J1112,
we report TESS-based pulsation frequencies for the first time. For the remaining targets, we present new and/or reprocessed TESS short- and ultra-short-cadence observations, and we derive updated frequency solutions for all of them. We then conduct a detailed asteroseismological analysis to constrain the stellar structure and fundamental parameters, in particular, the stellar mass and, importantly, the H-envelope mass, since asteroseismology remains the only method capable of estimating this quantity. To this end, we employ our grid of fully evolutionary models of LM He-core WDs that 
consistently allow for variations in the three relevant parameters at play  \citep[$T_{\rm eff}$, $M_\star$, and $M_{\rm H}$; see][]{2018A&A...614A..49C,2018A&A...620A.196C}. 
Guided by the H-envelope mass suggested by asteroseismology,  
we subsequently derive spectroscopic/photometric stellar mass estimates based on the atmospheric parameters available for each star.

This work is organized as follows. In Section~\ref{targets}, we describe the ELMV sample and the data reduction and frequency analysis based on TESS photometry. In Section~\ref{astero}, we carry out a detailed asteroseismological analysis of the sample to constrain the stellar mass and H-envelope mass. In Section~\ref{photo-spec_masses}, we present the spectroscopic/photometric mass estimates. Finally, in Section~\ref{summary}, we summarize our main results and present our conclusions.

\section{Target stars, TESS observations, and data reduction}
\label{targets}

In this Section, we describe the TESS observations and data reduction for the six ELMV targets included in this study. One of these objects is analyzed here for the first time using TESS data, while in most cases, we incorporate additional Sectors not previously considered in the literature. Some of these stars have been analyzed in earlier works \citep{2022MNRAS.511.1574R,2025ApJ...984..112R,2024A&A...684A..76B}; however, we homogeneously reprocess all available data and carry out an independent frequency extraction based on an extended dataset and a refined pre-whitening approach.

Our choice of these six stars was guided by pragmatic considerations: the availability of short- (120\,s) or ultra-short-cadence (20\,s) TESS data covering at least one sector, the presence of more than one independent pulsation mode with S/N above our adopted FAP threshold in preliminary frequency analyses, and the availability of atmospheric parameters. These conditions favor targets with relatively clean but non-trivial pulsation spectra, making them particularly suitable for this first TESS-based asteroseismological study of ELMVs.

Table~\ref{table:ELMV-sample} in Appendix \ref{appendix:properties} summarizes the properties of the six ELMV targets analyzed in this work. For J1112 \citep[the second discovered ELMV;][]{2013ApJ...765..102H}, we adopt spectroscopic parameters with 3D corrections following \citet{2015ApJ...809..148T}. The remaining five targets were reported by \citet{2022MNRAS.511.1574R}; for these stars, we adopt Gaia-based atmospheric parameters of $T_{\rm eff}$ and $\log g$ from \citet{2021MNRAS.508.3877G}.

We describe below the corresponding TESS observations, data reduction, and target-by-target frequency analysis. We used the signal prewhitening code \href{https://pyriod.readthedocs.io/en/latest/}{\texttt{Pyriod}} \citep{2022ascl.soft07007B} to detect and characterize significant pulsation periods. We adopted the amplitude thresholds recommended by \cite{2021AcA....71..113B}, corresponding to a 0.1\% false-alarm probability (FAP). Any adopted signals falling below this threshold are marked with an asterisk and discussed in the text. For the reported amplitudes and signal-to-noise ratios (S/N), we selected the highest value observed across the sectors. For each object, we analyzed all available sectors, prioritizing ultra-short-cadence data whenever it was available. All significant signals that we detect are below the Nyquist frequency for the 120-s short cadence data, and the best-fit parameters are insensitive to the data set used. We combined measurements of the same frequencies detected in multiple sectors using inverse-variance weighting.

\begin{figure*}[t!]
    \centering
    \includegraphics[width=\textwidth, keepaspectratio]{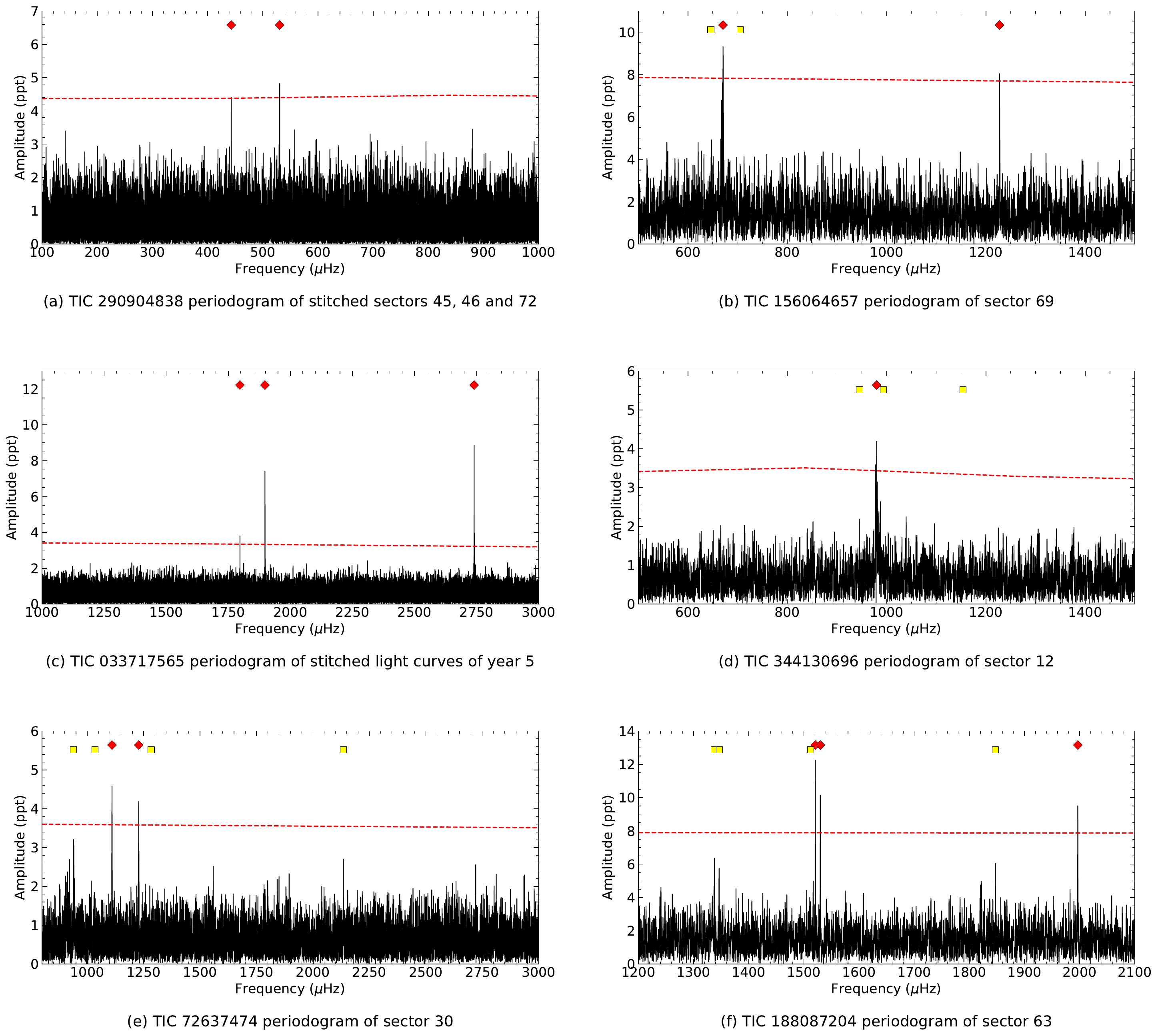}
    \caption{Representative amplitude spectra for each target, selected from the TESS sector (or stitched sectors) that most clearly show the pulsation signals. Significant frequencies detected in the displayed spectrum are marked with red diamonds. The dashed red line represents the significance threshold. Yellow squares highlight frequencies that were detected in the comprehensive analysis but are weak or not visible in this specific plot.}
    \label{fig:all_targets_periodograms}
\end{figure*}

\subsection{TIC~290904838 (J1112)} 
\label{data_j1112}

This star was first identified from high-speed photometric observations at the McDonald Observatory by \cite{2013ApJ...765..102H}. It exhibited seven pulsation periods spanning from $107$ to $2856\ $s (although the two shortest periods of $107$ and $134\ $s, suggested by \citeauthor{2013ApJ...765..102H} to be possible pressure($p$)-mode pulsations, still require confirmation). 

We present here the first analysis of TESS data for J1112. This target was observed in three short-cadence (SC) sectors: 45, 46, and 72. No single sector shows a peak above the $0.1\%$ FAP threshold. However, when the light curves are stitched across sectors, the signal-to-noise ratio improves, and two peaks exceed the threshold. The pulsation frequencies derived from our
analysis are listed in Table \ref{tab:frequencies_tic290904838}. 
Notably, two of the longest periods reported by \cite{2013ApJ...765..102H} are now confirmed by the TESS observations.
The amplitude spectrum for this target, based on the stitched light curves of Sectors 45, 46, and 72, is shown in Figure \ref{fig:all_targets_periodograms} (panel a).

\begin{table}
    \centering
      \caption{Frequencies, periods, amplitudes, S/N, and FAP values for TIC~290904838.}
       \resizebox{\columnwidth}{!}{
        \begin{tabular}{lccccc}
            \toprule
            ID & Frequency [$\mu\mathrm{Hz}$] & Period [s] & Amplitude [ppt] & S/N &FAP [\%] \\
            \midrule
            $f_0$ & $530.622 \pm 0.020$ & $1884.581 \pm 0.071$ & $5.2 \pm 1.0$ & 5.86  & $\num {7.4e-4}$ \\
            $f_1$ & $442.781 \pm 0.021$ & $2258.45 \pm 0.11$ & $5.0 \pm 0.9 $ & 5.23 &$\num{7.8e-2}$\\
            \bottomrule
        \end{tabular}
    }
    \label{tab:frequencies_tic290904838}
\end{table}

\subsection{TIC~156064657}
\label{data_tic156064657}

TESS observed this star in Sector 29 (SC) and Sectors 69 and 96 (both SC and ultra-short cadence, USC). Following the cadence-selection criterion described above, USC data were used whenever available. The pulsation frequencies derived from our analysis are listed in Table \ref{tab:frequencies_tic156064657}. Figure \ref{fig:all_targets_periodograms} (panel b) displays the periodogram for Sector 69.
We identified a signal in Sector 29 that falls below the significance threshold but is likely a pulsation frequency, labeled as $f_3^*$. Accepting this signal would increase the FAP to $3.44\%$ (see Appendix \ref{appendix:FAP}). In our model fitting, we consider subsets of measured periods both including and excluding $f_3^*$.

For comparison, \citet{2022MNRAS.511.1574R} analyzed only Sector 29 and reported two periodicities, which we recover ($f_1$ and $f_3^*$). By incorporating additional sectors (and USC data), our analysis yields a more complete pulsation set, adding two significant signals ($f_0$ and $f_2$).

\begin{table}
    \centering
    \caption{Same as Table~\ref{tab:frequencies_tic290904838}, but for TIC~156064657}
    \resizebox{\columnwidth}{!}{
        \begin{tabular}{lccccc}
            \toprule
            ID & Frequency [$\mu\mathrm{Hz}$] & Period [s] & Amplitude [ppt] & S/N & FAP [\%] \\
            \midrule
            $f_0$ & $1227.509 \pm 0.029$ & $814.658 \pm 0.019$ & $8.1 \pm 1.2$ &  5.57 & \num{1.3e-2}\\
            $f_1$ & $705.197 \pm 0.022$ & $1418.043 \pm 0.044$ & $13.4 \pm 1.2$ & 9.79 & $\ll \num{1e-6}$\\
            $f_2$ & $670.574 \pm 0.024$ & $1491.260 \pm 0.053$ & $9.5 \pm 1.3$ & 6.41 & $\num{1.6e-5}$\\
            $f_3^*$ &  $645.957 \pm 0.049$ &  $1548.11 \pm 0.12$ & $6.1 \pm 1.2$ & 4.47 & $3.44$\\
            \bottomrule
        \end{tabular}
    }
    \label{tab:frequencies_tic156064657}
\end{table}

\subsection{TIC~33717565}
\label{data_tic33717565}

This WD was observed with SC in 24 sectors (17 also have USC data). Given the large number of light curves, we stitched the data by observing year to reduce frequency uncertainties; the comparatively small frequency errors for this target result from this yearly stitching. For the stitched light curves, the 0.1\% FAP detection threshold follows the method described by \cite{2021AcA....71..113B}. The resulting frequencies are listed in Table \ref{tab:frequencies_tic33717565}. The periodogram for Year 5 is presented in Figure \ref{fig:all_targets_periodograms} (panel c).

It is worth noting that \cite{2025ApJ...984..112R} \citep[see also][]{2022MNRAS.511.1574R} carried out a frequency analysis for TIC~33717565 using the same sectors and reported a larger set of seven frequencies; however, the corresponding periodogram is not shown. Our analysis confirms two of these signals ($f_0$ and $f_1$) and identifies one additional pulsation frequency ($f_2$).

\begin{table}
    \centering
       \caption{Same as Table~\ref{tab:frequencies_tic290904838}, but for TIC~33717565}
    \resizebox{\columnwidth}{!}{%
        \begin{tabular}{lccccc}
            \toprule
            ID & Frequency [$\mu\mathrm{Hz}$] & Period [s] & Amplitude [ppt] & S/N & FAP [\%] \\
            \midrule
            $f_0$ & $2740.323 \pm 0.001$ & $364.920 \pm 0.001$ & $19.3 \pm 1.8$ & 13.53 & $\ll \num{1e-6}$\\
            $f_1$ & $1897.633 \pm 0.001$ & $526.972 \pm 0.001$ & $11.2 \pm 1.2$ & 9.13 &  $\ll \num{1e-6}$\\
            $f_2$ & $1796.906 \pm 0.004$ & $556.512 \pm 0.001$ & $9.8 \pm 1.2$ & 6.09 & $\num{2.2e-3}$\\
            \bottomrule
        \end{tabular}
    }
     \label{tab:frequencies_tic33717565}
\end{table}

\subsection{TIC~344130696} 
\label{data_tic344130696}

This WD was observed in Sectors 12, 13, 39, 66, 93, and 94. In most sectors, both SC and USC data are available; we prioritized the USC when present. In Sector 93, the data in both cadences are noisy, and no peak above the 0.1\% FAP threshold was detected.
Two peaks detected in separate sectors appear to trace the same mode ($979.972\pm 0.031$ and $981.788 \pm 0.021\ \mu\mathrm{Hz}$); we averaged them and report the result as $f_2$. The complete list of frequencies is provided in Table \ref{tab:pulsations_clean}.
A representative amplitude spectrum from Sector 12 is plotted in Figure \ref{fig:all_targets_periodograms} (panel d).

\cite{2022MNRAS.511.1574R} presented a similar analysis, but limited to data from Sectors 12, 13, and 39. We recover their two reported modes ($f_2$ and $f_3$). With the expanded dataset, we identify two additional significant signals ($f_0$ and $f_1$).

\begin{table}
    \centering
    \caption{Same as Table~\ref{tab:frequencies_tic290904838}, but for TIC 344130696} 
    \resizebox{\columnwidth}{!}{
        \begin{tabular}{lccccc}
            \toprule
            ID & Frequency [$\mu\mathrm{Hz}$] & Period [s] & Amplitude [ppt] & S/N & FAP [\%] \\
            \midrule
            $f_0$ & $1153.637 \pm 0.033$ & $866.824 \pm 0.025$ & $5.83 \pm 0.80$ & 5.70 & $ \num{5.1e-3}$ \\
            $f_1$ & $993.569 \pm 0.038$ & $1006.473 \pm 0.038$ & $4.10 \pm 0.70$ & 5.26 & $ \num{2.1e-2}$ \\
            $f_2$ & $981.217 \pm 0.017$ & $1019.143 \pm 0.018$ & $5.11 \pm 0.55$ & 7.78 & $\ll \num{1e-6}$ \\
            $f_3$ & $945.612 \pm 0.018$ & $1057.516 \pm 0.020$ & $5.69 \pm 0.48$ & 10.03 & $\ll \num{1e-6}$ \\
            \bottomrule
        \end{tabular}
    }

    \label{tab:pulsations_clean}
\end{table}

\subsection{TIC~72637474}
\label{data_tic72637474}

TESS observed this object in Sector 3 (SC) and Sectors 30 and 97 (SC and USC). We detected five significant peaks above the 0.1\% FAP threshold. 
We also identified a sixth, lower-amplitude peak that falls below this threshold, which we regard as a plausible candidate signal. Including this peak would require relaxing the FAP threshold to 0.857\% (see Appendix \ref{appendix:FAP}). The effect of including this signal will be tested in our asteroseismological model fitting.  This signal is labeled $f_5^*$ in Table \ref{tab:frequencies_tic72637474}.
Figure \ref{fig:all_targets_periodograms} (panel e) shows the periodogram for Sector 30, highlighting the significant frequencies.

For comparison, \cite{2022MNRAS.511.1574R} analyzed data from Sectors 3 and 30 and reported only three periods, all of which we recover ($f_2$, $f_3$, and $f_4$). 
We also detect an additional signal, $f_0$, already visible in the Sector 30 data, although it was not reported by \cite{2022MNRAS.511.1574R}. With the expanded dataset, we identify one further significant signal ($f_1$), plus one low-amplitude candidate ($f_5^*$).

\begin{table}
    \centering
    \caption{Same as Table~\ref{tab:frequencies_tic290904838}, but for TIC~72637474.}  
    \resizebox{\columnwidth}{!}{
        \begin{tabular}{lccccc}
            \toprule
            ID & Frequency [$\mu\mathrm{Hz}$] & Period [s] & Amplitude [ppt] & S/N & FAP [\%] \\            
            \midrule
            $f_0$ & $2135.362 \pm 0.017$ & $468.305 \pm 0.004$ & $3.14 \pm 0.46$ & 5.83 & $ \num{1.9e-3}$ \\
            $f_1$ & $1283.145 \pm 0.011$ & $779.335 \pm 0.007$ & $4.9 \pm 1.4 $ & 9.14 & $\ll \num{1e-6}$ \\
            $f_2$ & $1228.012 \pm 0.031$ & $814.324 \pm 0.021$ & $4.15 \pm 0.56$ & 6.13 &  $ \num{1.7e-4}$ \\
            $f_3$ & $1109.743 \pm 0.028$ & $901.110 \pm 0.023$ & $4.54 \pm 0.56$ & 6.68 & $ \num{1.6e-6}$ \\
            $f_4$ & $1034.186 \pm 0.042$ & $966.944 \pm 0.039$ & $5.02 \pm 0.73$ & 6.12& $ \num{3.1e-5}$ \\
            $f_5^*$ & $938.952 \pm 0.043$ & $1065.017 \pm 0.049$ & $3.20 \pm 0.54$ & 4.71 &$ \num{0.857}$ \\
            \bottomrule
        \end{tabular}
    }
    \label{tab:frequencies_tic72637474}
\end{table}

\subsection{TIC~188087204}
\label{data_tic188087204}

This WD was observed in Sectors 63 and 90 (SC and USC) and in Sector 36 (SC). 
We detected seven signals, which are listed in Table~\ref{table:188087204}. Five of them are significant peaks ($f_0$, $f_2$, $f_3$, $f_4$, and $f_5$), while two additional sub-threshold signals ($f_1^*$ and $f_6^*$) are detected in Sectors 63 and 90 and stand out above the surrounding noise in both datasets.

These two signals were not identified through a blind search: each was first noticed in the sector where it has the higher amplitude and then recovered at the same frequency in the other sector. We therefore assess their significance in this second, lower-amplitude sector using a rank-based criterion. If $n$ peaks exceed the amplitude of the candidate within the
searched band, the probability that noise produces a matching peak of at least that rank is approximately $P \simeq 2n/N$, where $N$ is the cadence-dependent number of independent frequency bins and the factor of two accounts for the
two-sided frequency-matching tolerance. For both $f_1^*$ (1847~$\mu$Hz) and $f_6^*$ (1337~$\mu$Hz), four peaks exceed the candidate in the relevant sector, giving $P \simeq 6.9\times10^{-3}\%$ in each case, well below our nominal
detection threshold. This supports treating $f_1^*$ and $f_6^*$ as plausible pulsation modes.

The pulsation spectrum for this star is illustrated in Figure~\ref{fig:all_targets_periodograms} (panel f), using data from Sector 63. Three primary pulsation frequencies, \(f_2\), \(f_3\), and \(f_4\), exhibit a clear pattern of nearly uniform spacing. Averaging the consecutive separations yields \(\Delta f = 8.878 \pm 0.029~\mu\mathrm{Hz}\).  This structure can be naturally interpreted as a rotationally split triplet, with \(f_3\) as the central (\(m = 0\)) component. An independent pair, \(f_5\) and \(f_6^*\), shows a consistent spacing of \(8.960 \pm 0.055~\mu\mathrm{Hz}\), reinforcing the same pattern, and thus, suggesting the possible presence of an additional, incomplete triplet. On this basis, we treat \(f_3\) as a robust \(m = 0\) candidate, while \(f_5\) and \(f_6^*\) are considered alternative \(m = 0\) identifications for the putative second triplet.

Based on Sector 36 alone, \citet{2022MNRAS.511.1574R} reported five pulsation frequencies all of which we recover ($f_0$, $f_1^*$, $f_3$, $f_4$, and $f_5$). Both that work and \citet{2024A&A...684A..76B} noted evidence of a possible triplet near $1521\ \mu$Hz (which we identify as $f_2$, $f_3$, and $f_4$). Our expanded multi-sector analysis not only confirms that structure, but also reveals evidence for an additional incomplete pair of frequencies (\(f_5\) and \(f_6^*\)) with consistent spacing.

\begin{table}
    \centering
        \caption{Same as Table~\ref{tab:frequencies_tic290904838}, but for TIC~188087204}
    \resizebox{\columnwidth}{!}{%
        \begin{tabular}{lccccc}
            \toprule
         ID & Frequency [$\mu\mathrm{Hz}$] & Period [s] & Amplitude [ppt] & S/N & FAP [\%] \\
            \midrule
            $f_0$ & $1996.577 \pm 0.023$ & $500.857 \pm 0.006$ & $9.5 \pm 1.3$ & 6.42 & $ \num{1.5e-5}$\\
            $f_1^*$ & $1847.156 \pm 0.055$ & $541.373 \pm 0.016$ & $6.0 \pm 1.2$ & 4.43  & $ \num{6.9e-3}$ \\      $f_2$ & $1529.837 \pm 0.032$ & $653.664 \pm 0.014$ & $10.1 \pm 1.2$ & 6.78 & $\ll \num{1e-6}$ \\      
            $f_3$ & $1520.877 \pm 0.016$ & $657.515 \pm 0.007$ & $15.2 \pm 1.6$ & 8.11 & $\ll \num{1e-6}$ \\        $f_4$ & $1512.081 \pm 0.043$ & $661.340 \pm 0.019$ & $12.7 \pm 1.6$ & 5.21 &  $ \num{2.9e-2}$\\    
            $f_5$ & $1346.797 \pm 0.021$ & $742.502 \pm 0.012$ & $14 \pm 3$  & 8.25 & $\ll \num{1e-6}$\\
            $f_6^*$ & $1337.837 \pm 0.051$ & $747.475 \pm 0.028$ & $6.4 \pm 1.2$ & 4.27 & $ \num{6.9e-3}$\\
            \bottomrule
        \end{tabular}
    }
    \label{table:188087204}
\end{table}

\section{Asteroseismological modeling}  
\label{astero}

The pulsational analysis in this work is based on a set of stellar models computed with the {\tt LPCODE} evolution code \citep{2005A&A...435..631A, 2009A&A...502..207A,2013A&A...557A..19A,2015A&A...576A...9A}, which provides a self-consistent description of the internal structure and chemical stratification by following the complete evolutionary history of LM He-core WDs from the ZAMS. The models of LM WDs we employ in this work include both thick (canonical)  and thin H envelopes \citep[see][for details about input physics and evolutionary calculations]{2013A&A...557A..19A,2018A&A...620A.196C}. Adiabatic pulsation periods for non-radial dipole ($\ell = 1$) and quadrupole ($\ell = 2$) $g$-modes were computed employing the adiabatic version of the {\tt LP-PUL} pulsation code \citep[see][for details]{2006A&A...454..863C}. These periods correspond to the central $m=0$ components of any rotational multiplets.

We determine the stellar mass and H-envelope thickness, among other relevant quantities, by identifying the model whose theoretical pulsation periods best reproduce the observed ones \citep{2019A&ARv..27....7C}. The quality of the match between the theoretical periods ($\Pi_k^{\rm T}$) and observed ones ($\Pi_i^{\rm O}$) is quantified through the merit function: 

\begin{equation}
\label{sigma2}
\sigma^2(M_{\star},  T_{\rm   eff}, M_{\rm H})=   \frac{1}{n} \sum_{i=1}^{n}   \min \left[(\Pi_i^{\rm   O}-   \Pi_k^{\rm  T})^2 \right], 
\end{equation}

\noindent where $n$ is the number of observed periods and $k$ denotes the radial order of the theoretical pulsation mode. The LM WD model that yields the minimum value of $\sigma^2$, if it exists, is adopted as the best-fit model. We calculate $\sigma^2=\sigma^2(M_{\star}, T_{\rm eff}, M_{\rm H})$ for the following discrete stellar masses: $0.1554$, $0.1612$, $0.1650$, $0.1706$, $0.1762$, $0.1805$, $0.1869$, $0.1921$, $0.2025$,  $0.2390$, $0.2707$,  $0.3205$, $0.3624$, and $0.4352\ M_{\sun}$. For the H-envelope mass, we consider a wide range, $-5.8 \lesssim \log(M_{\rm H}/M_{\star}) \lesssim -1.7$, depending on the stellar mass; the detailed set of H-envelope thicknesses is described in \citet{2018A&A...620A.196C}. Along each evolutionary sequence, the available models cover $13000 \gtrsim T_{\rm eff} \gtrsim 6000\ $K. Since $T_{\rm eff}$ is determined by the evolutionary calculations, the local step size in $T_{\rm eff}$ depends on the sequence and evolutionary stage; around the adopted solutions, it ranges from $10$ to $50\ $K.

Analogously to \cite{2018A&A...620A.196C}, we start our analysis assuming that all the observed periods correspond to $g$ modes associated with $\ell= 1$, and considering the set of observed periods of each target star to compute the quality function given by Eq.~(\ref{sigma2}). Next, we assume that the observed periods correspond to a mixture of $\ell=1$ and $\ell=2$ $g$ modes. Since we usually do not find suitable solutions when assuming $\ell=1$ alone, we display here the cases in which a combination of $\ell=1$ and $\ell=2$ $g$ modes is assumed. In all cases, $m=0$ is assumed for the observed modes, which is uncertain for most of these signals where the complete rotationally split multiplet is not detected ($f_3$ from TIC~188087204 is an exception). In Figures~\ref{fig:j1112} to \ref{fig:tic188087204-s4} we show, for each ELMV target, the projection of $\log\sigma$ on the $T_{\rm eff}$–$M_{\star}$ plane. 
Each point in these maps corresponds to the value of the H-envelope mass that minimizes $\sigma$ for the given $(T_{\rm eff}, M_{\star})$. The color scale is chosen such that smaller values of $\log\sigma$ denote better fits, and the displayed ranges in $T_{\rm eff}$ have been adjusted to highlight the region of interest around the atmospheric parameters. If a single, well-defined minimum is present for a given star, we adopt the corresponding model as the asteroseismological solution. When multiple possible solutions exist --- an expected behavior given that pulsating LM WDs generally show only a few independent modes ---, we use the available atmospheric information (see Sect.~\ref{targets}) to guide the selection of a representative solution and/or to define a range of acceptable models. In the $M_{\star}$-$T_{\rm eff}$ maps, these atmospheric values are shown as $1 \sigma$ error bars.  For the Gaia-based cases, the plotted bars provide a compact visualization in this plane, although the underlying constraints on $M_{\star}$ and $T_{\rm eff}$ are correlated. The atmospheric parameters are therefore used only as guidance when evaluating competing minima. A circle marks the adopted representative asteroseismological solution, when one can be identified. 

Table~\ref{table:ELMV-masses} summarizes the main results of the asteroseismological analysis of this work, indicating $M^{\rm astero}_{\star}$, $\log(M_{\rm H}/M_{\star})$, and $T^{\rm astero}_{\rm eff}$\footnote{These quantities correspond to selected models in our evolutionary sequences, and the number of quoted digits should not be interpreted as a formal estimate of their precision.}. It also includes the atmospheric parameters ($T_{\rm eff}$ and $\log g$), either spectroscopic or photometric, and the corresponding spectroscopic or photometric masses derived later in Section~\ref{photo-spec_masses}. Next, we describe the individual asteroseismological results.

\begin{table*} 
\centering
\caption{Main characteristics of the adopted asteroseismological models for the ELMVs analyzed in this work. }

\begin{tabular}{ccccccc}
\hline
\noalign{\smallskip}
Star &  $T_{\rm eff} [K]$& $\log(g)$ & $M^{\rm spec}_{\star}/M^{\rm phot}_{\star}[M_{\sun}]$ &  $M^{\rm astero}_{\star} [M_{\sun}]$ & $\log(M_{\rm H}/M_{\star})$ & $T^{\rm astero}_{\rm eff} [K]$\\
\hline
\noalign{\smallskip}
J1112$^*$ & $9240$ & $6.170$ & $0.140$  & $0.1706$ & $-5.30$& $8922$\\
TIC~156064657  & $10194$ & $7.295$&  $0.327$ & $0.3624$ & $-3.10$  & $10202$\\
TIC~33717565   & $10676$ & $7.639$ & $(0.424,0.445)$  & $(0.1706,0.3624)$ & $(-5.54,-2.50)$ & $(10660,11200)$\\ 
TIC~344130696$^*$  & $10829$ & $7.177$ & $0.300$   & $0.1869$ & $-2.37$& $10632$\\
TIC~72637474  & $10214 $ & $7.209$ & $0.292$   & $0.2707$ & $-3.67$& $10126$\\
TIC~188087204  & $10052$ & $7.583$ & $(0.415,0.423)$ & $(0.3624,0.4352)$ & $(-5.79,-3.62)$& $(9729,9820)$ \\

\noalign{\smallskip}
\noalign{\smallskip}
\hline
\noalign{\smallskip}
\end{tabular}
\tablefoot{Column 1 lists the star name. Columns 2 and 3 give the adopted atmospheric parameters, effective temperature and surface gravity, either spectroscopic (J1112) or Gaia-based (the remaining five stars). Column 4 lists the spectroscopic mass for J1112 and the photometric mass for the other stars. Column 5 gives the asteroseismological stellar mass obtained from our period-to-period fits, while columns 6 and 7 list, respectively, the corresponding H-envelope mass and effective temperature of the adopted asteroseismological model. A range of values is provided when only a range of acceptable solutions was found. $^*$ Tentative asteroseismological solution (see text for details).}
\label{table:ELMV-masses}
\end{table*}

\subsection{TIC~290904838 (J1112)}
\begin{figure}
\centering
 \includegraphics[clip,width=1.0\linewidth]{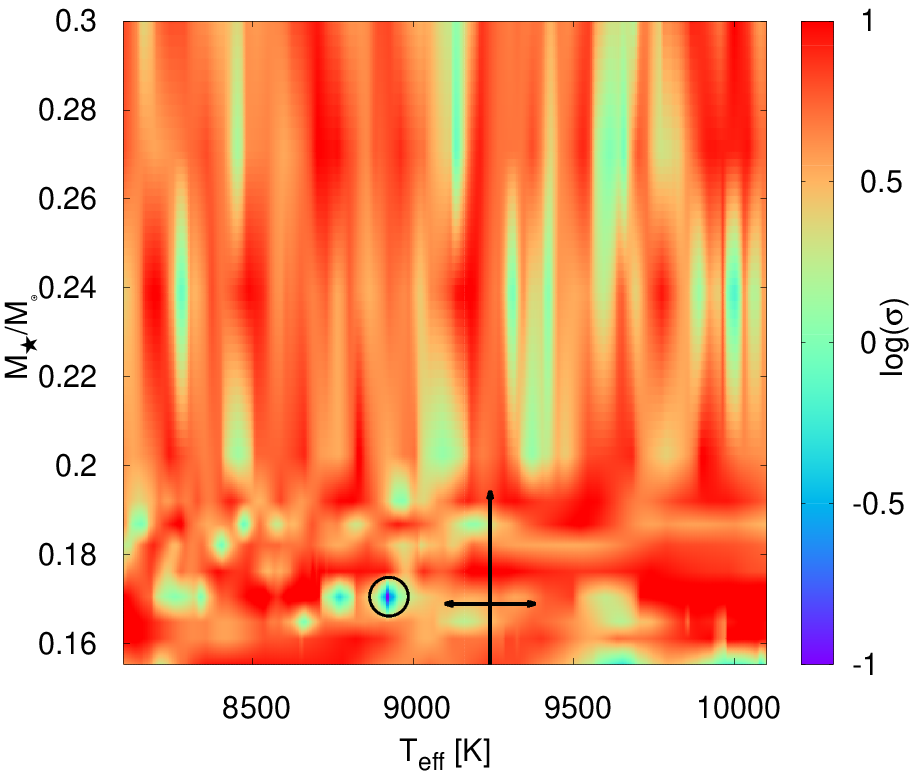}
	 \caption{Projection on the $T_{\rm eff}$ versus $M_{\star}$ plane of the logarithm of the average period residual ($\sigma$) for J1112, assuming that the observed periods are associated with $\ell= 1, 2$. For each $M_\star$, the value shown corresponds to the H–envelope mass that yields the lowest value of the quality function among the explored sequences. The error bars indicate the $1\sigma$ atmospheric estimates in the $M_{\star}$-$T_{\rm eff}$ plane. The circle marks the adopted solution.}  
	\label{fig:j1112}
\end{figure}

Considering the two independent periods observed for this star ($\sim 1885$ and $2258\ $s) and reported in Sect.~\ref{data_j1112}, we performed period-to-period fits assuming that both correspond to a mixture of $g$ modes with $\ell= 1$ and $2$, as described above. The results are displayed in Fig.~\ref{fig:j1112}. A narrow region of low $\sigma$ is found along the $0.1706\ M_{\odot}$ sequence, at a stellar mass very close to the spectroscopic estimate ($M_{\star} \sim 0.169 \ M_{\odot}$), but at effective temperatures around $8920\ $K, somewhat lower than the spectroscopic estimate ($T_{\rm eff} \sim 9240 \ $K). The absolute minimum of the quality function (i.e. the smallest $\sigma$) occurs at $T_{\rm eff}=8922\ $K, $M_{\star}=0.1706\ M_{\odot}$, and $\log(M_{\rm H}/M_{\star})=-5.30$, corresponding to a very thin-H envelope, and yields an average period residual of $\sigma \sim  0.03\ $s. However, consecutive models along the same $(M_{\star},M_{\rm H})$ sequence in the range $T_{\rm eff}\sim 8916-8928\ $K provide comparably good fits, with $\sigma \sim 0.2\ $s. Typical changes in the theoretical periods between adjacent grid points are of order $0.1$-$0.4\ $s. Residuals much smaller than this, such as $\sigma\sim 0.03\ $s, are therefore below the intrinsic resolution of our grid. We therefore regard the consecutive models within this small region as equivalent solutions for our purposes and retain the solution in the range $T_{\rm eff}\sim 8916-8928\ $K, with $M_{\star}=0.1706\ M_{\odot}$ and $\log(M_{\rm H}/M_{\star})=-5.30$, as our preferred asteroseismological solution for J1112. This inference must nevertheless be regarded with caution, as it is based on only two independent observed periods.

\subsection{TIC~156064657}

Given the set of periods reported in Sect.~\ref{data_tic156064657} for this WD ($\sim 815$, $1418$, $1491$, and $1548\ $s), we considered two subsets of periods for our period-to-period fits. Set S1 comprises the three highest-significance periods, $\sim 815$, $1418$, and $1491\ $s, which satisfy our original FAP threshold, while set S2 includes all four, adding the $\sim 1548\ $s signal, which becomes acceptable when we relax the FAP criterion. 

When using only the three periods in S1 and inspecting the results in our full grid, the absolute minimum of the quality function occurs at an effective temperature 
$T_{\rm eff}\sim 9300\ $K, well below the $T_{\rm eff}$ of the star ($\sim 10\,195\ $K). If we focus on a narrower region of the $M_{\star}$-$T_{\rm eff}$ plane, we find a local minimum near $T_{\rm eff}= 10\,190\ $K with $M_{\star}= 0.3624\ M_{\odot}$ and $\log(M_{\rm H}/M_{\star})= -3.10$ (canonical envelope), yielding an average period residual of $\sigma \sim 1.5\ $s. However, additional solutions 
with similar fit quality are also present, so the corresponding $\log\sigma$ map does not reveal a clearly preferred solution and is therefore not shown here. 

\begin{figure}
\centering
 \includegraphics[clip,width=1.0\linewidth]{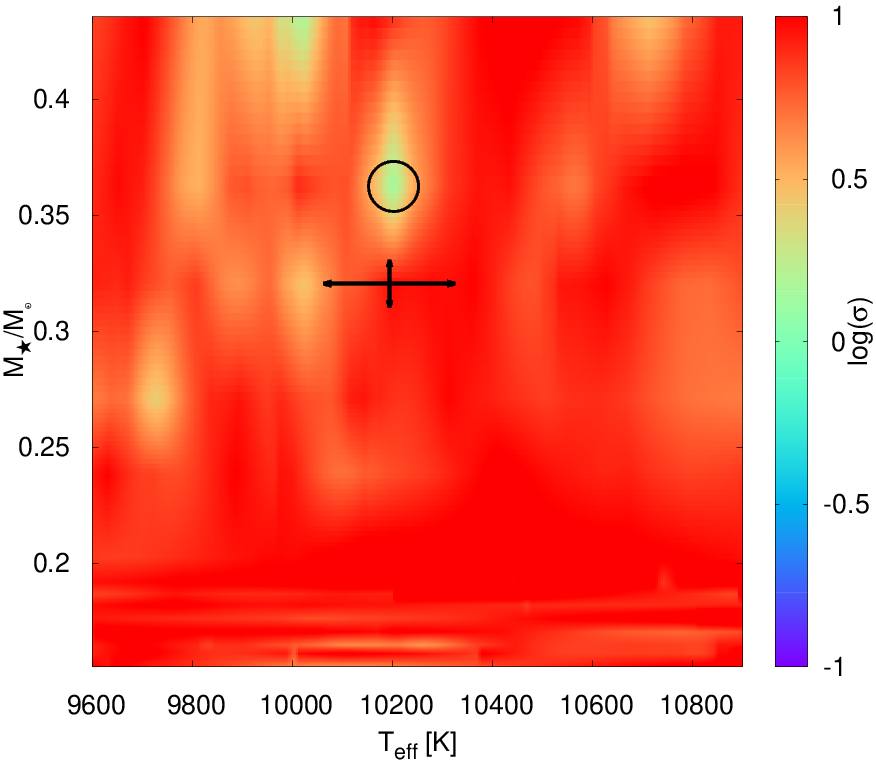}
	 \caption{Same as Fig.~\ref{fig:j1112}, but for TIC~156064657 considering the periods from the set S2.}  	\label{fig:tic4657-s2}
\end{figure}

Repeating the procedure with S2 (i.e., including the additional $1548\ $s period), yields the results shown in Fig.~\ref{fig:tic4657-s2}. In this case, the global minimum of the quality function --- i.e., the best fit over the entire explored parameter space --- is found for a model very close to the local minimum identified in the S1 case. This model has $T_{\rm eff}= 10202\ $K, $M_{\star}= 0.3624\ M_{\odot}$, and $\log(M_{\rm H}/M_{\star})= -3.10$ (canonical envelope) and is characterized by $\sigma \sim 1.4\ $s.
Although this fit results from the addition of the $\sim 1548\ $s period (obtained after relaxing the FAP criterion), its inclusion improves the overall period match, and the solution is better constrained, as it removes alternative solutions present in the case of S1. We therefore adopt this model as representative of TIC~156064657.

\subsection{TIC~33717565}
\begin{figure}
\centering
 \includegraphics[clip,width=1.0\linewidth]{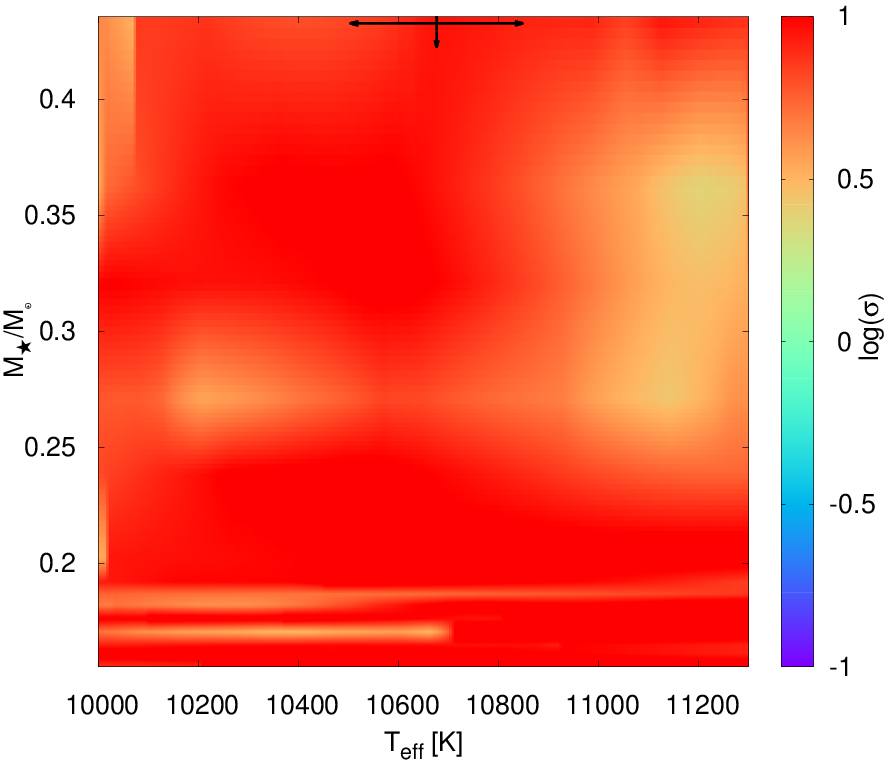}
	 \caption{Same as Fig.~\ref{fig:j1112}, but for TIC~33717565.}  
	\label{fig:tic33717565}
\end{figure}

Similarly to the previous cases, we performed period-to-period fits for TIC~33717565 using the three periods identified in Sect.~\ref{data_tic33717565} (i.e., $\sim 365$, $527$, and $557\ $s). Over the full grid, the global minimum of the quality function is found at a relatively low $T_{\rm eff}$ ($\sim 9600\ $K), well below the atmospheric estimate for the star ($T_{\rm eff} \sim 10680\ $K). In a narrower range, Fig.~\ref{fig:tic33717565} reveals multiple possible solutions. Three models are particularly noteworthy. The first has $T_{\rm eff}= 11200\ $K, $M_{\star}= 0.3624\ M_{\odot}$, $\log(M_{\rm H}/M_{\star})= -3.10$ (canonical envelope). The second, also at a relatively high effective temperature, has $T_{\rm eff}= 11140\ $K, $M_{\star}= 0.2707\ M_{\odot}$, $\log(M_{\rm H}/M_{\star})= -5.54$ (very thin envelope). The third has $T_{\rm eff}= 10658\ $K, $M_{\star}= 0.1706\ M_{\odot}$, $\log(M_{\rm H}/M_{\star})= -2.50$ (thin envelope). All three yield comparable average period residuals of a few seconds. None reproduces the atmospheric estimates for the star particularly well: the third lies closest to the $T_{\rm eff}$ estimate, whereas the first two provide slightly better period fits, but all three are found at substantially lower masses than the Gaia-based estimate ($\sim 0.433\ M_{\odot}$). We therefore do not identify a uniquely preferred asteroseismological solution for TIC~33717565. 
Given this degeneracy, we can only conclude that acceptable solutions span stellar masses in the range $0.1706 - 0.3624\ M_{\odot}$, $T_{\rm eff}$ between $10660 - 11200\ $K, and a very poorly constrained H envelope mass between $\sim 3 \times 10^{-6} M_{\star}$ and $3 \times 10^{-3} M_{\star}$.

\subsection{TIC~344130696}
\begin{figure}
\centering
 \includegraphics[clip,width=1.0\linewidth]{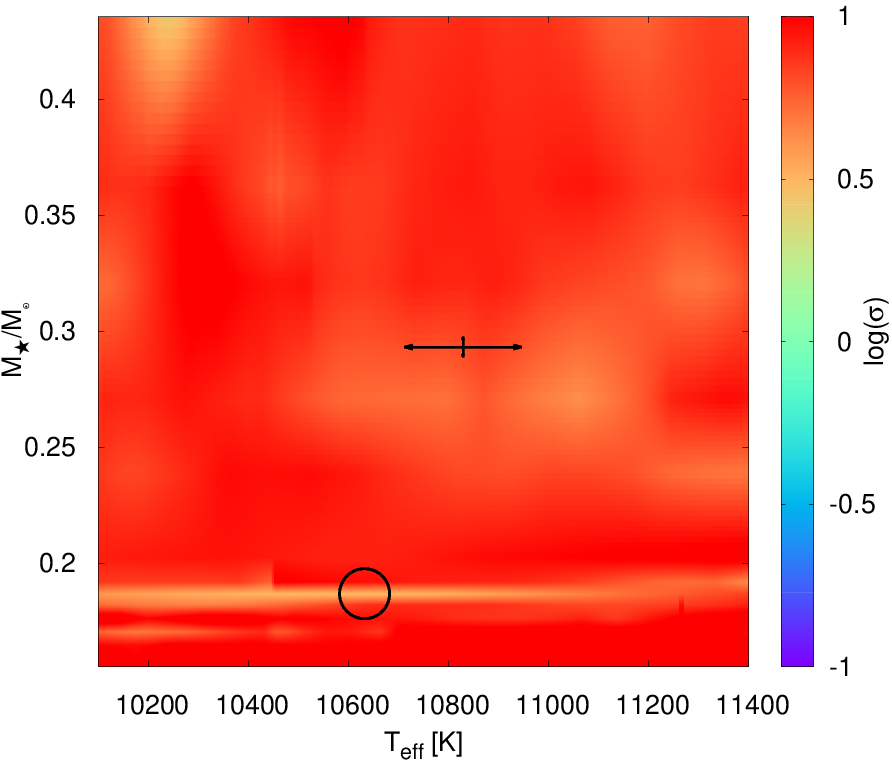}
	 \caption{Same as Fig.~\ref{fig:j1112}, but for TIC~344130696.}  
	\label{fig:tic344130696}
\end{figure}

We performed period-to-period fits for TIC~344130696 employing the four periods determined in Sect.~\ref{data_tic344130696} (i.e., $\sim 867$, $1006$, $1019$, and $1058\ $s). Over the full grid, the best period fit is located at a very low $T_{\rm eff}$ ($\sim 9680\ $K), well below the estimate for the star ($T_{\rm eff} \sim 10830\ $K). Within the narrower range of $T_{\rm eff}$, Fig.~\ref{fig:tic344130696} reveals the existence of multiple possible solutions. Two models are of particular interest. 
The first has $T_{\rm eff}= 10246\ $K, $M_{\star}= 0.4352\ M_{\odot}$, $\log(M_{\rm H}/M_{\star})= -3.21$ (canonical envelope), and $\sigma\sim 2.7\ $s, while the second has $T_{\rm eff}= 10632\ $K, $M_{\star}= 0.1869\ M_{\odot}$, $\log(M_H/M_{\star})= -2.37$ (canonical envelope), and $\sigma\sim 3.0\ $s. Both provide comparable average period residuals. The higher-mass model yields the slightly smaller $\sigma$, but the lower-mass model lies closer to the effective temperature and stellar mass estimates of the star.  
We therefore regard TIC~344130696 as having only a tentative asteroseismological solution and retain the lower-mass model as the most plausible one. 

\subsection{TIC~72637474}
As in the case of TIC~156064657, for TIC~72637474, we performed two separate period-to-period fits based on two different subsets of the detected periods. Set S1 comprises the five periods listed in Table~\ref{tab:frequencies_tic72637474} which lie above the $0.1\%$ FAP threshold ($\sim 468$, $779$, $814$, $901$, $967\ $s; see Sect.~\ref{data_tic72637474}). Set S2 extends S1 by including the additional signal $f_5^*$ at $\sim 1065\ $s, which becomes acceptable only when the FAP threshold is relaxed to $0.857\%$.

\begin{figure}
\centering
 \includegraphics[clip,width=1.0\linewidth]{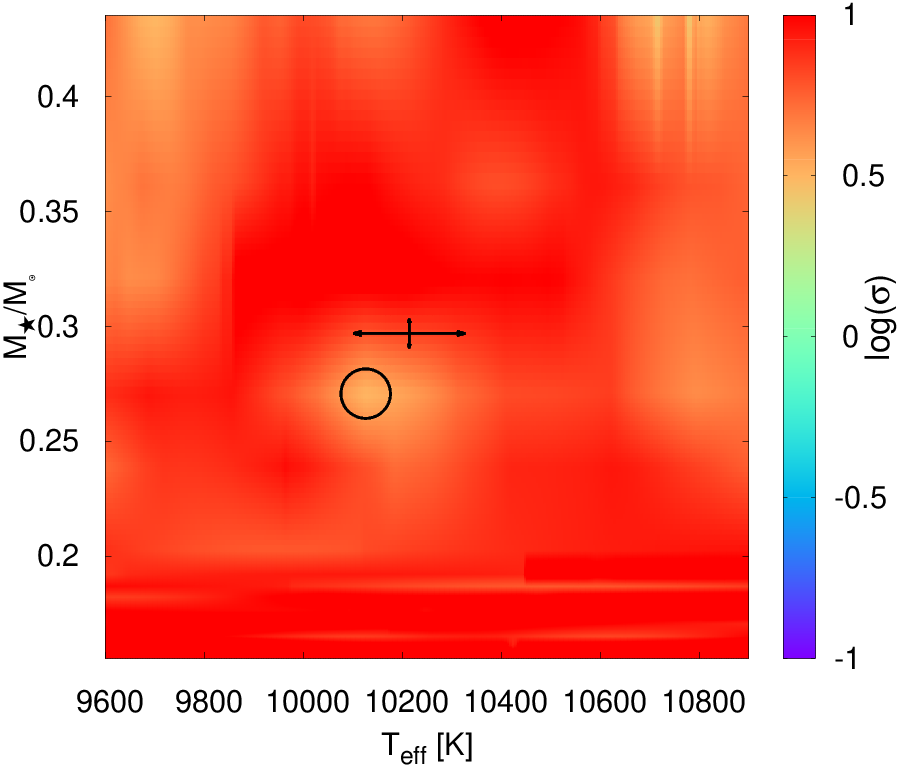}
	 \caption{Same as Fig.~\ref{fig:j1112}, but for TIC~72637474, considering the periods from the set S1.}  
	\label{fig:tic72637474-s1}
\end{figure}

Using the five periods in S1, we find that the absolute minimum of the quality function over the explored parameter space is reached at $T_{\rm eff} \sim 12930\ $K, much hotter than the atmospheric estimate for the star ($T_{\rm eff} \sim 10215\ $K). Within the narrower interval shown in Fig.~\ref{fig:tic72637474-s1}, some possible solutions with comparable quality ($\sigma \sim 3\ s$) are visible. Most of them, however, lie at effective temperatures and masses clearly discrepant with the atmospheric estimates. By contrast, the model with $T_{\rm eff}= 10126\ $K, $M_{\star}= 0.2707\ M_{\odot}$, and $\log(M_H/M_{\star})= -3.67$ (thin envelope) lies closest to the atmospheric parameters. 

When repeating the analysis with S2 (i.e., including the additional $\sim 1065\ $s period), the overall period-fit quality deteriorates. We therefore do not show the S2 results here and adopt the S1 solution with $T_{\rm eff}= 10126\ $K, $M_{\star}= 0.2707\ M_{\odot}$, and $\log(M_H/M_{\star})= -3.67$ as the representative model for TIC~72637474.

\subsection{TIC~188087204}
For TIC~188087204, we performed period-to-period fits using four different subsets of the detected periods, hereafter S1--S4, each reflecting different assumptions regarding period significance and mode identification (see Sect.~\ref{data_tic188087204}). Set S1 comprises the three periods at $\sim 501$, $658$, and $743\ $s, corresponding to the frequencies $f_0$, $f_3$, and $f_5$ listed in Table~\ref{table:188087204}, and assumes that $f_5$ provides the $m=0$ component of the putative second triplet, with $f_6^*$ excluded. Set S2 instead adopts $f_6^*$ as the $m=0$ component and comprises the periods at $\sim 501$, $658$, and $747\ $s ($f_0$, $f_3$, and $f_6^*$). Sets S3 and S4 extend S1 and S2, respectively, by including the additional sub-threshold candidate $f_1^*$ at $\sim 541\ $s, thus yielding S3 with $\sim 501$, $541$, $658$, and $743\ $s, and S4 with $\sim 501$, $541$, $658$, and $747\ $s. By construction, S1 uses only periods above the nominal FAP threshold, while S2–S4 treat $f_1^*$ and/or $f_6^*$ as plausible, though less secure, modes (Sect.~\ref{data_tic188087204}).

\begin{figure}
\centering
 \includegraphics[clip,width=1.0\linewidth]{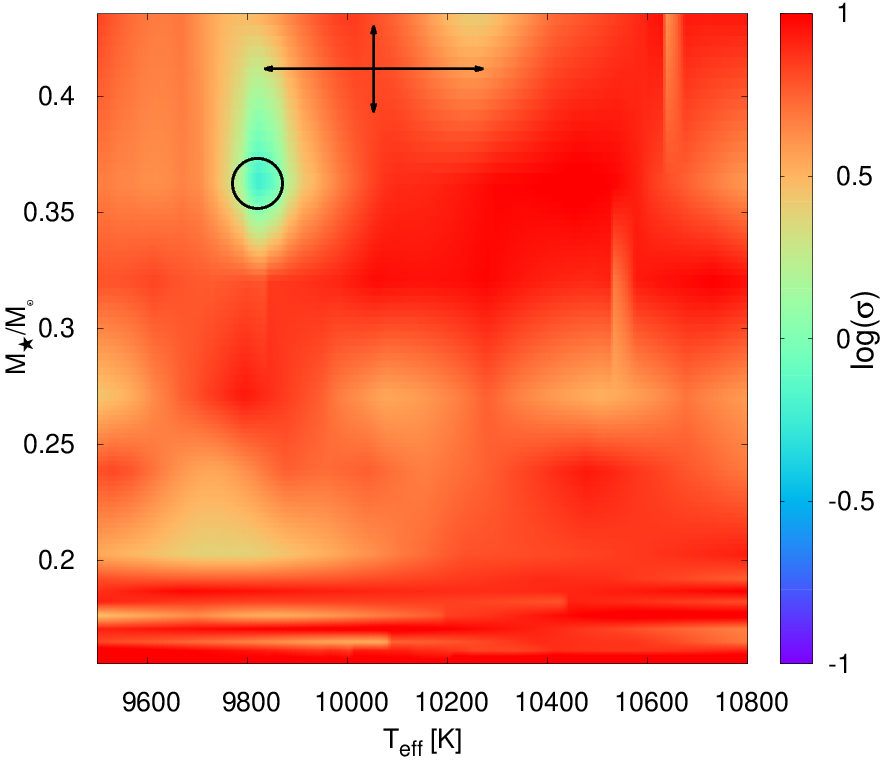}
	 \caption{Same as Fig.~\ref{fig:j1112}, but for TIC~188087204 considering the periods from set S1, which includes only frequencies above the nominal FAP threshold and adopts $f_5$ as the $m=0$ candidate of the putative second triplet.}  
	\label{fig:tic188087204-s1}
\end{figure}

Using the three periods in S1 ($\sim 501$, $658$, and $743\ $s), the best period fit over the full grid occurs at $T_{\rm eff}= 9354\ $K, 
well below the atmospheric estimate for the star ($T_{\rm eff} \sim 10050\ $K), and we therefore discard this solution. Within the narrower region shown in Fig.~\ref{fig:tic188087204-s1}, however,
a viable solution is found at $T_{\rm eff}= 9820\ $K, $M_{\star}= 0.3624\ M_{\odot}$, and $\log(M_H/M_{\star})= -3.62$ (thin envelope), with $\sigma \sim 0.5\ $s. This solution is slightly cooler and somewhat less massive than the Gaia-based estimates. 

Repeating the procedure with the S2 set ($\sim 501$, $658$, and $747\ $s), the overall quality of the fits 
deteriorates. The best period fit is found at $T_{\rm eff} \sim 7545\ $K, well below the atmospheric estimate for this star, and we therefore discard that solution. Once again, restricting the analyzed region, the best local solution is found at $T_{\rm eff}= 9729\ $K, $M_{\star}= 0.4352\ M_{\odot}$, and $\log(M_H/M_{\star})= -4.32$ (thin envelope), yielding $\sigma \sim 1\ $s. Although this solution lies closer in mass to the atmospheric estimate than the S1 one, it is cooler and yields a somewhat poorer period match. Since the S2 subset does not provide a more compelling alternative than the other subsets, we do not display its results here.
The S3 set ($\sim 501$, $541$, $658$, and $743\ $s) produces numerous solutions scattered throughout the parameter space, all with significantly larger $\sigma$ than those obtained from S1. The absence of a well-defined minimum and the overall low quality of the fits make this configuration uninformative for our purposes, and we do not display its map either.

\begin{figure}
\centering
 \includegraphics[clip,width=1.0\linewidth]{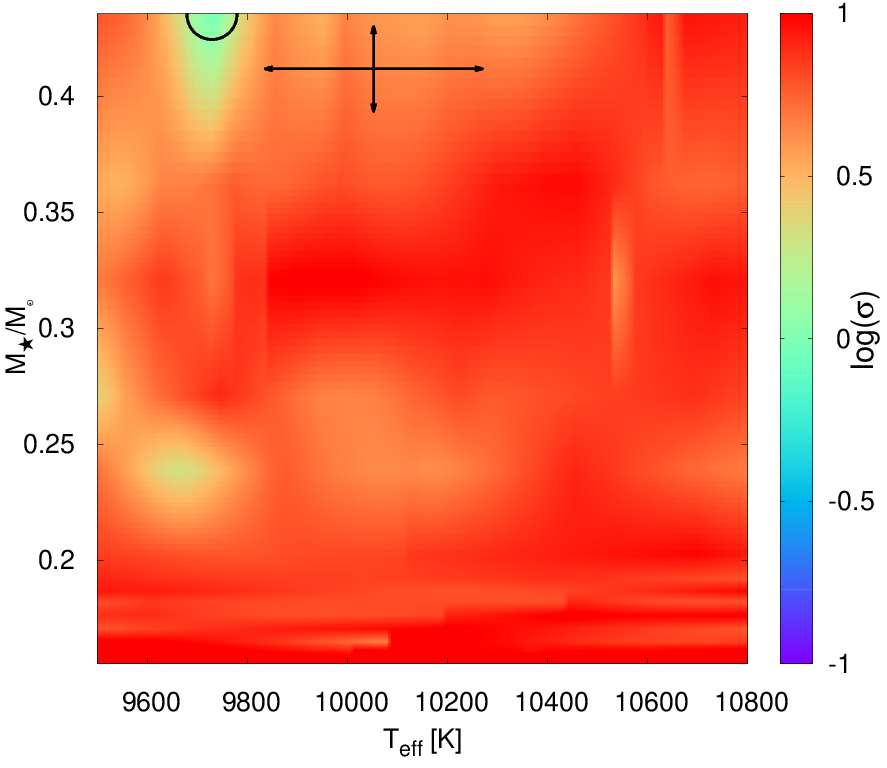}
	 \caption{Same as Fig.~\ref{fig:tic188087204-s1}, but for the periods from set S4, which includes the additional sub-threshold candidate $f_1^*$ and adopts $f_6^*$ instead of $f_5$ as the $m=0$ candidate of the putative second triplet.}  
	\label{fig:tic188087204-s4}
\end{figure}

The S4 set ($\sim 501$, $541$, $658$, and $747\ $s) yields the most relevant alternative to S1. 
As shown in Fig.~\ref{fig:tic188087204-s4}, the global minimum of the quality function over the entire explored grid, is found at $T_{\rm eff}= 9729\ $K, $M_{\star}= 0.4352\ M_{\odot}$, and $\log(M_H/M_{\star})= -5.79$ (very-thin envelope), with $\sigma \sim 0.9\ $s. Relative to the S1 solution, this model is cooler but lies closer in mass to the atmospheric estimate.
Also, the strength of its period match --- being the best-fit solution for this set --- makes it a relevant solution. A secondary solution, but with larger residuals ($\sigma \sim 2.1\ $s), appears at $T_{\rm eff}= 9658\ $K, $M_{\star}= 0.2390\ M_{\odot}$, and $\log(M_H/M_{\star})= -5.15$ (very-thin envelope).

Overall, we can only conclude that the asteroseismological solutions for TIC~188087204 favor stellar masses between $0.3624 - 0.4352\ M_{\odot}$, with $T_{\rm eff}$ between $9729 - 9820\ $K, and H-envelope mass roughly in the range $2 \times 10^{-6} - 2 \times 10^{-4} M_{\star}$. Among the considered sets, S1 provides the best compromise between period-fit quality and mode identification. Nonetheless, because an acceptable solution also arises from S4, we adopt the above intervals as our conservative asteroseismological inference for this star.

\section{Spectroscopic/photometric masses}
\label{photo-spec_masses}

We estimated the spectroscopic/photometric stellar masses of our ELMV sample by interpolating their atmospheric parameters ($T_{\rm eff}$, $\log g$) on the evolutionary sequences of \cite{2013A&A...557A..19A} and \cite{2018A&A...620A.196C} in the $\log g - \log T_{\rm eff}$ plane (the “Kiel diagram”). For J1112, these parameters are spectroscopic (\citealt{2013ApJ...765..102H}, corrected for 3D effects following \citealt{2015ApJ...809..148T}), whereas for the remaining five stars, they are Gaia–based values \citep{2021MNRAS.508.3877G}. As discussed above, the stellar mass of an LM WD depends on the envelope thickness ---thinner envelopes imply higher surface gravities at fixed mass \citep{2018A&A...614A..49C,2025A&A...699A.280A}. To account for this effect, we employed the two-track families cited above, thus spanning from canonical to very thin H envelopes.

As shown in Sect.~\ref{astero} and summarized in Table~\ref{table:ELMV-masses}, the preferred asteroseismological solutions for the stars in our sample span a broad range of H-envelope thicknesses. J1112 is best reproduced, albeit tentatively, by models with very thin H envelopes; TIC~72637474 is consistent with thin H envelopes; TIC~33717565 allows solutions from canonical to very thin envelopes; TIC~188087204 is best reproduced by models ranging from thin to very thin envelopes; and the remaining two stars (TIC~156064657 and TIC~344130696) are consistent with canonical envelopes. 
Guided by the envelope-thickness regimes suggested by the asteroseismological fits, we computed for each star its spectroscopic/photometric mass by interpolating on the corresponding evolutionary tracks. For TIC~33717565 and TIC~188087204, for which more than one envelope-thickness regime is possible, we evaluated the consistent cases and report the interval spanned by the two values. In cases where the asteroseismological solution is tentative, the resulting spectroscopic/photometric mass should be interpreted accordingly.

\begin{figure}
\centering
 \includegraphics[width=\columnwidth,trim=0.9cm 0.5cm 0.21cm 0.2cm,clip]{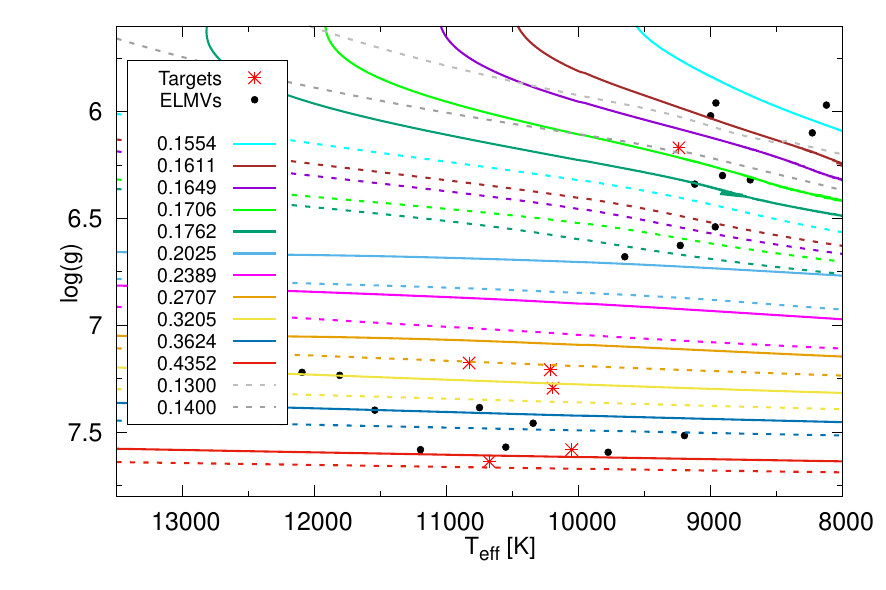}
	\caption{Kiel diagram for the six ELMVs (red crosses) analyzed in this work. Evolutionary tracks for canonical \citep[solid;][]{2013A&A...557A..19A} and very thin \citep[dashed;][]{2018A&A...620A.196C} H-envelope masses are included, as well as two artificial evolutionary sequences (with $0.130$ and $0.140\ M_{\odot}$) with very thin H-envelopes. Also displayed are other ELMV targets (black dots). All masses are given in units of $M_{\odot}$.}
	\label{fig:01}
\end{figure}

The positions of our target stars in the Kiel diagram are shown in Figure~\ref{fig:01}, together with the evolutionary sequences employed \citep[solid and dashed lines;][respectively]{2013A&A...557A..19A,2018A&A...620A.196C}. For clarity, we only display evolutionary tracks corresponding to the canonical and very thin H-envelope limits of our model grid; intermediate envelope masses ---although relevant for some of our asteroseismological solutions--- are not shown in order to avoid overcrowding the figure. We also included two evolutionary sequences with $0.130$ and $0.140\ M_{\odot}$ and very thin H envelopes, extracted from \cite{2018A&A...614A..49C}, to cover the domain of low surface gravities of our grid. The resulting spectroscopic/photometric masses, $M^{\rm spec}_{\star}$/$M^{\rm phot}_{\star}$, are listed in the fourth column of Table~\ref{table:ELMV-masses}. 

Both the spectroscopic and photometric masses inherit uncertainties from the adopted atmospheric parameters. A detailed analysis of the spectroscopic masses will be addressed in a subsequent paper \cite[see e.g.,][and references therein, for a recent discussion of the uncertainties affecting mass estimates in pulsating DA WDs]{2024A&A...691A.194C}. For the five stars with Gaia-based parameters, the photometric masses include an additional layer of model dependence that is specific to this work. Their $(T_{\rm eff}, \log g)$ values are taken from \cite{2021MNRAS.508.3877G}, who fit Gaia photometry and parallaxes with model atmospheres combined with the canonical LM He-core WD sequences of \cite{2001MNRAS.325..607S}. For our $M^{\rm phot}_{\star}$ estimates, we instead interpolate these same parameters on the canonical grids of \cite{2013A&A...557A..19A} and on their thin-envelope extensions \citep{2018A&A...620A.196C}. As a rough estimate of the associated systematics, we find that the choice of canonical He-core grid changes $M_{\rm phot}$ by only $\sim 0.006\ M_\odot$, while adopting very thin instead of canonical H envelopes can shift $M_{\rm phot}$ by up to $\sim 0.026\ M_\odot$. This latter effect is relevant only for
TIC~33717565 and TIC~188087204, the two stars for which our asteroseismology already yields a range of acceptable solutions rather than a single mass value, and, to a lesser extent, TIC~72637474, since it fits a thin-H envelope model. Nonetheless, a more homogeneous and self-consistent set of photometric masses, computed by fitting multi-band photometry and parallax with our full grid of He-core evolutionary models (including thin-H envelopes when required), will be presented in a forthcoming paper.

\section{Summary and conclusions}
\label{summary}

In this work, we presented the first homogeneous TESS-based and detailed asteroseismological study of a sample of six pulsating LM He-core WDs. For J1112, we analyzed TESS data for the first time, and detected two periods that confirm two of the longest-period modes previously reported with ground observations \citep{2013ApJ...765..102H}. For the remaining stars, we provide new or updated frequency solutions with respect to previous studies.
In particular, for TIC~188087204, we confirm a previously reported triplet and uncover evidence for a second, incomplete multiplet with consistent rotational splitting.

We performed detailed period-to-period fits using our fully evolutionary models and the observed periods for each star. 
These fits yield well-constrained asteroseismological solutions for three stars, a representative but more tentative solution for a fourth target, and constrained ranges for the remaining two. Our results suggest that the inferred H-envelope masses span a broad range (from canonical to very thin), covering the range of envelope structures explored by available evolutionary calculations \citep{2013A&A...557A..19A,2025A&A...699A.280A,2018A&A...614A..49C}. 
Using the atmospheric parameters available for each star, we derived spectroscopic/photometric mass estimates by interpolating on evolutionary tracks consistent with the envelope thickness regime suggested by asteroseismology. In most cases, the resulting stellar masses are broadly compatible with the corresponding asteroseismological values. 
The main exceptions are TIC~33717565, for which the photometric masses remain higher than the asteroseismological estimates over the limiting envelope cases considered, and TIC~344130696, whose asteroseismological solution remains more tentative. Finally, we note that unresolved companions may affect Gaia-based photometric parameters and derived photometric masses. Although we find broad agreement between photometric and asteroseismological masses for most targets, spectroscopic follow-up will be valuable to assess this possibility on a star-by-star basis.

Despite this general level of agreement among the different estimates of the stellar mass, it should be kept in mind that the present asteroseismological inferences remain subject to important limitations.
While the WD models employed in this analysis are among the most detailed and comprehensive available ---particularly regarding stellar mass, effective temperature, and H-envelope thickness--- the asteroseismological fits for several targets remain poorly constrained due to the limited number of detected periods and the lack of secure mode identification.
In addition, the model comparison assumes that the observed modes are the central $m=0$ components of rotational multiplets, although this is uncertain for most of these signals where we only detect one component of the expected multiplet structure. This may be a significant source of systematic error, as the study of TESS observations of the ELMV pulsator GD\,278 demonstrated that rotational splittings can exceed the spacing between adjacent radial overtones \citep{2021ApJ...922..220L}. These caveats are particularly relevant when interpreting the apparent diversity in H-envelope thicknesses across the sample. Thus, the inferred H-envelope masses are in some cases better regarded as indicative values than as unique determinations. In particular, for J1112, where only two independent periods are currently available, the preferred very-thin-envelope solution should be considered provisional, and we refrain from drawing firm conclusions about the formation channel or possible flash history of this star on that basis alone. Additional independent modes and/or complementary constraints will be required to assess the robustness of this result.

Finally, we compare our results with previous asteroseismological studies. For J1112, using the same fully evolutionary models adopted here, \cite{2018A&A...620A.196C} derived a representative model with $M_{\star}=0.1612\ M_{\sun}$ from the five longest periods reported by \citet{2013ApJ...765..102H}, whereas we obtain $M_{\star}=0.1706\ M_{\sun}$ from the two modes confirmed with TESS. 
This modest discrepancy is likely due to the different set of fitted periods. For the remaining stars, our preferred solutions differ substantially from those reported by \cite{2022MNRAS.511.1574R}, who modeled all targets as ZZ Ceti stars (i.e., pulsating DA WDs with CO cores), and thus employed CO–core evolutionary models. This naturally leads to higher inferred stellar masses: $0.493$, $0.632$, and $0.542\ M_{\sun}$ for TIC~156064657, TIC~344130696, and TIC~72637474, respectively, compared with our $0.3624$, $0.1869$, and $0.2707\ M_{\sun}$; and $0.609$ and $0.493\ M_{\sun}$ for TIC~33717565 and TIC~188087204, compared with our ranges of $0.1706$-$0.3624$ and $0.3624$-$0.4352\ M_{\sun}$. These discrepancies, particularly for the least massive cases, are therefore not unexpected and reflect the different evolutionary framework. In addition, our updated frequency solutions imply revised fitted period sets, which may also lead to a different preferred asteroseismological model.

The analysis presented here represents a first step towards a systematic asteroseismological characterization of pulsating LM WDs with TESS data. In this work, we have focused on a subset of ELMVs for which TESS already reveals relatively clean frequency spectra and for which basic atmospheric parameters are available, making them particularly suitable for a first homogeneous analysis. Extending the same methodology to the remaining ELMV candidates observed by TESS, many of which are expected to show more complex or lower-S/N pulsation spectra, will require a dedicated analysis of their light curves and frequency content and is beyond the scope of this first study. Future space-based photometric missions are also expected to expand the sample of candidates and, in many cases, provide richer mode sets that enable more robust mode identification and stronger constraints on theoretical models (e.g., \citealt{2025arXiv251119196U}).

\begin{acknowledgements}
The authors warmly thank the referee for their comments and suggestions, 
which helped to improve this work. We thank the Asociación Argentina de Astronomía for supporting the publication costs of this article.
 M. U. gratefully acknowledges funding from the Research Foundation Flanders (FWO) by means of a junior postdoctoral fellowship (grant agreement No. 1247624N). 
 NH and KJB are supported by NASA through grant No.\ 80NSSC25K0122 of the TESS Cycle 7 General Investigator Program. KJB is also supported by the National Science Foundation under grant No.\ AST-2406917.
 We thank S.~O.~Kepler for helpful comments on the manuscript while it was under TASC review.
 This paper includes data collected with the TESS mission, obtained from the MAST data archive at the Space Telescope Science Institute (STScI). Funding for the TESS mission is provided by the NASA Explorer Program.
 This research has made use of NASA's Astrophysics Data System Bibliographic Services, and the SIMBAD and VizieR databases, operated at CDS, Strasbourg, France.

\end{acknowledgements}   
  
\bibliographystyle{aa}
\bibliography{biblio}  

\clearpage
\begin{appendix}

\section{Lists of selected objects}
\label{appendix:properties}

Here we provide basic information about the objects analyzed in this paper.

\begin{table}[!h]
\centering
\rotatebox{90}{%
\begin{minipage}{0.82\textheight}
\caption{Main properties of the sample of ELMVs analyzed in this work. Column (a) is the star's name, column (b) indicates its Gaia DR3 ID, columns (c) and (d) show the equatorial coordinates. Column (e) displays the DR3  Gaia apparent magnitude. Columns (f), (g), and (h) list the effective temperature, gravity, and stellar mass (spectroscopic for J1112, Gaia-based for the rest), respectively, while column (i) is the reference.}
\label{table:ELMV-sample}
\centering
\setlength{\tabcolsep}{3pt}
\begin{tabular}{lcccccccc}
\hline
Star & GAIA ID & RA    &   DEC   &    $G$  &   $T_{\rm eff}$ & $\log(g)$ & $M_*$ & Reference   \\     
      &        & (deg) &  (deg)  &   (mag) &       [K]       & [cgs] & [$M_{\odot}$] & \\
\hline                    
(a)   &  (b)   &  (c)   & (d)    & (e)     & (f)     & (g)        & (h) & (i)     \\  \hline          
J1112& 3963587822967516032& $11:12:15.80$ & $+11:17:45.1$& $16.34$ & $ 9240 \pm 140 $ &  $6.17 \pm 0.060$ & $0.169 \pm 0.025$ & H13, T15\\ 
TIC~156064657 & 4974784825671467648  & $00:37:23.75$ & $-48:21:55.9$  & $16.57$ & $ 10194 \pm 132 $ & $7.295 \pm 0.034$ & $0.320 \pm 0.010$& GF21\\
TIC~33717565 & 4627855367706529152   & $04:05:36.39$ & $-76:28:28.1$ & $16.50$ & $10676 \pm 173  $ & $7.639 \pm 0.031$ & $0.433 \pm 0.010$& GF21\\
TIC~344130696 & 6365271657299575680  & $18:37:08.30$ &$-76:59:05.9$& $15.39$ & $10829 \pm 116  $ & $7.177 \pm 0.018$ & $0.293 \pm 0.004$ &GF21\\
TIC~72637474 & 5020319141229055360   & $02:08:07.86$ & $-29:31:38.8$ & $15.90$ & $10214 \pm 113  $ & $7.209 \pm 0.025$ & $0.297 \pm 0.006$&GF21\\
TIC~188087204 & 5470271185153118208  & $10:46:27.80$ & $-25:12:15.8$  & $16.83$ & $10052 \pm 218  $ & $7.583 \pm 0.055$ & $0.412 \pm 0.019$ &GF21 \\
\hline
\end{tabular}
\vspace{2mm}
\parbox{\linewidth}{\small
References. H13: \cite{2013ApJ...765..102H}; T15: \cite{2015ApJ...809..148T}; GF21: \cite{2021MNRAS.508.3877G}.}
\end{minipage}}
\end{table}

\normalsize
\newpage

\section{Methodology for re-computing FAP for sub-threshold peaks}
\label{appendix:FAP}

When a candidate peak falls below the nominal 0.1\% detection threshold, we reassess its significance by computing a new false-alarm probability (FAP) using the analytic model of \cite{2021AcA....71..113B}. Their Eq.~(5) gives the percentile \(u_p\) of the median-standardized amplitude-spectrum maximum \(U\) as a function of the number of data points \(N\) and FAP \(p\). In practice, we measure \(u_p\) for a candidate by normalizing the peak amplitude to the spectrum’s local/median noise level (consistent with our periodogram workflow), and then infer \(p\) by inverting Eq.~(5). This inversion is closed-form (a Gumbel-distribution), so evaluation is straightforward once \(u_p\) and \(N\) are specified.

\paragraph{Inputs.}
(1) Cadence-informed \(N\). Following \cite{2021AcA....71..113B}, we adopt the per-sector data-point counts \(N=116{,}640\) for 20\,s (ultra-short) cadence and \(N=19{,}440\) for 120\,s (short) cadence.\\
(2) Search bandwidth. If the test is confined to a subrange of the spectrum (e.g., a white-dwarf \(g\)-mode band), we replace \(N\) by an effective \(N_{\rm eff}=f\,N\), where \(f\) is the tested frequency range as a fraction of \((0,\nu_{\rm Nyq}]\). This aligns the calculation with the actual number of independent frequencies being interrogated. \\
(3) Measured \(u_p\). For each candidate, we compute the percentile-like height \(u_p\) from the median standardized amplitude spectrum of the (residual) time series used to test that peak.

\paragraph{Computation.}
With \(u_p\) and \(N\) (or \(N_{\rm eff}\)) in hand, we invert Eq.~(5) of \cite{2021AcA....71..113B} to obtain
\[
p \;=\; 1 - \exp\!\Bigl(-\exp\!\bigl(\tfrac{1.05\,\ln N - (u_p/1.201)^2}{1.04}\bigr)\Bigr).
\]
We report this \(p\) as the recalculated FAP for that candidate. 

\end{appendix}

\end{document}